\documentclass[12pt]{article}

\usepackage[dvips]{graphicx}  

\usepackage[english]{babel} 

\begin{document}

\begin{center}
{\Large\bf  Resolving Spacecraft Earth-Flyby  Anomalies with   Measured Light Speed Anisotropy  \rule{0pt}{13pt}}\par

\bigskip
Reginald T. Cahill \\ 
{\small\it  School of Chemistry, Physics and Earth Sciences, Flinders University,
Adelaide 5001, Australia\rule{0pt}{15pt}}\\
\raisebox{+1pt}{\footnotesize E-mail: Reg.Cahill@flinders.edu.au}\par

Published:  Progress in Physics  {\bf 3}, 9-15,2008.
\bigskip

{\small\parbox{11cm}{%
Doppler shift observations  of spacecraft, such as Galileo, NEAR, Cassini, Rosetta and MESSENGER in earth flybys,  have all revealed unexplained  speed `anomalies' - that the doppler-shift determined speeds are inconsistent with expected speeds.  Here it is shown that these speed anomalies are not real and are actually the result of using an incorrect relationship between the observed doppler shift and the speed of the spacecraft - a relationship  based on the assumption that the speed of light is isotropic in all frames, {\it viz} invariant. Taking account of the repeatedly measured light-speed anisotropy the anomalies are resolved {\it ab initio}.    The Pioneer 10/11 anomalies are discussed, but not resolved.
The spacecraft observations demonstrate again that the speed of light is not invariant, and is isotropic only with respect to a dynamical 3-space. The existing doppler shift data  also offers a resource to characterise a new form of gravitational waves, the dynamical 3-space turbulence, that has also been  detected by other techniques.  The Einstein spacetime formalism uses a special definition of space and time coordinates that mandates light speed invariance for all observers, but which is easily misunderstood and misapplied. \rule[0pt]{0pt}{0pt}}}\medskip
\end{center}{%

\setcounter{section}{0}
\setcounter{equation}{0}
\setcounter{figure}{0}
\setcounter{table}{0}

\markboth{R.T.  Cahill. Resolving Spacecraft Earth-Flyby  Anomalies with   Measured Light Speed Anisotropy }{\thepage}
\markright{R.T.  Cahill. Resolving Spacecraft Earth-Flyby   Anomalies with   Measured Light Speed Anisotropy  }

\section{Introduction}
Planetary probe spacecraft (SC) have their speeds increased,  in the heliocentric frame of reference, by a close flyby of the earth, and other planets.  However in the earth frame of reference there should be no change in the asymptotic speeds after an earth flyby, assuming the validity of Newtonian gravity, at least in these circumstances. However doppler shift observations  of spacecraft, such as Galileo, NEAR, Cassini, Rosetta and MESSENGER in earth flybys,  have all revealed unexplained  speed `anomalies' - that the doppler-shift determined speeds are inconsistent with expected speeds
  \cite{And2008,ref1,ref3,ref4,ref5,ref6}.  Here it is shown that these speed anomalies are not real and are actually the result of using an incorrect relationship between the observed doppler shift and the speed of the spacecraft - a relationship  based on the assumption that the speed of light is isotropic in all frames, {\it viz} invariant. Taking account of the repeatedly measured light-speed anisotropy the anomalies are resolved {\it ab initio}.   

 The speed of light anisotropy has been detected  in at least 11 experiments \cite{MM,Miller,Illingworth,Joos, Jaseja,Torr,DeWitte,CahillCoax,CahillOF1,Munera,CahillOF2}, beginning with the Michelson-Morley 1887 experiment \cite{MM}. The interferometer observations and experimental techniques were first understood in 2002  when the Special Relativity effects and the presence of gas were used to calibrate the Michelson interferometer in gas-mode; in vacuum mode the Michelson interferometer cannot respond to  light speed anisotropy \cite{MMCK,MMC}, as confirmed in vacuum resonant cavity experiments, a modern version of the vacuum-mode Michelson interferometer \cite{cavities}. So far three different experimental techniques have given consistent results: gas-mode Michelson interferometers \cite{MM,Miller,Illingworth,Joos, Jaseja,Munera},  coaxial cable RF speed measurements  \cite{Torr,DeWitte,CahillCoax}, and optical-fiber Michelson interferometers \cite{CahillOF1,CahillOF2}.  This light speed anisotropy reveals the existence of a dynamical 3-space, with the speed of light being invariant only wrt that 3-space, and anisotropic according to observers in motion relative to that ontologically real frame of reference - such a motion being conventionally known as ``absolute motion", a notion thought to have been rendered inappropriate by the early experiments, particularly the Michelson-  Morley experiment.  However that experiment was never null - they reported a speed of at least 8km/s \cite{MM} using Newtonian physics for the calibration. A proper calibration of the Michelson-Morley apparatus gives a light speed anisotropy of at least 300km/s. The spacecraft doppler shift anomalies   are shown herein  to give  another technique  that may be used to measure the anisotropy of the speed of light, and give results consistent with previous detections.
 
 \begin{figure}[t]
\vspace{27mm}
\hspace{60mm}
\setlength{\unitlength}{1.5mm}
\hspace{0mm}\begin{picture}(0,0)
\thicklines
\put(6.5,5){\circle{20.0}}
\put(3,4.5){\bf Earth}
\qbezier(-5,6)(3,14)(14,16)
\put(14,16.1){\line(5,1){15}}

\put(-5,6){\line(-1,-1){10}}
\put(26,18.5){\vector(-4,-1){2}}
\put(25,14){\bf V}

\put(-10.9,0){\vector(-1,-1){2}}
\put(-12,-6){\bf V}

\end{picture}
\vspace{10mm}
\caption{\small{ Spacecraft (SC) earth flyby trajectory, with initial and final asymptotic velocity ${\bf V}$, differing only by direction. The doppler shift is determined from Fig.\ref{fig:doppler} and (\ref{eqn:frequency}). Assuming, as conventionally done,  that the speed of light is invariant in converting measured doppler shifts to deduced speeds, leads to the so-called flyby anomaly,  namely that the incoming and outgoing asymptotic speeds appear to be differ, by  $\Delta V_\infty$. However this effect is yet another way to observe the 3-space velocity vector, as well as 3-space wave effects, with the speed of light being $c$ and isotropic only wrt this structured and dynamical 3-space. The flyby  anomalies demonstrate, yet again,  that the invariance of the speed of light is merely a definitional aspect of the Einstein spacetime formalism, and  is not based upon observations.  A {\it neo}-Lorentzian 3-space {\it and}  time formalism is more physically appropriate.}}

 \label{fig:flyby}
\end{figure}
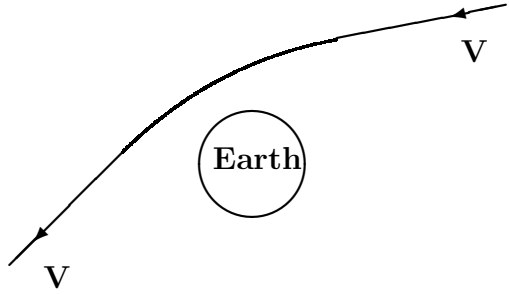

 The numerous light speed anisotropy experiments have also revealed turbulence in the velocity of the 3-space relative to the earth. This turbulence amounts to the detection of sub-mHz gravitational waves - which are present in the Michelson and Morley 1887 data, as discussed in \cite{Review}, and also present in the  Miller data   \cite{Miller,Book} also using a gas-mode Michelson interferometer, and by Torr and Kolen \cite{Torr}, DeWitte \cite{DeWitte}  and Cahill \cite{CahillCoax}  measuring RF speeds  in coaxial cables, and by Cahill \cite{CahillOF1}  and Cahill  and Stokes \cite{CahillOF2}  using an optical-fiber interferometer. 
The existing doppler shift data  also offers a resource to characterise this new form of gravitational waves.

The repeated detection of the anisotropy of the speed of light  is not in conflict with the results and consequences of Special Relativity (SR), although at face value it appears to be in conflict with Einstein's 1905 postulate that the speed of light is an invariant in vacuum.  However this  contradiction is more apparent than real, for one needs to realise that the space and time coordinates  used in the standard SR Einstein formalism are constructed to make the speed of light invariant wrt those special coordinates. To achieve that observers in relative motion must then relate their space and time coordinates  by a Lorentz transformation that mixes space and time coordinates - but this is only an artifact of this formalism.  Of course in the SR formalism one of the frames of reference could have always been designated as the observable   one. Such an ontologically real frame of reference,  only in which the speed of light is isotropic,  has been detected for over 120 years.  The usual literal interpretation of the 1905 Einstein postulate, {\it viz} that "the speed of light in vacuum is invariant", is actually experimentally shown to be false.  

There has been a long debate over whether the Lorentz 3-space {\it and} time interpretation or the Einstein spacetime interpretation of observed SR effects is  preferable or indeed even experimentally distinguishable.  
 What has been discovered in recent years is that a dynamical structured 3-space exists, so confirming the Lorentz interpretation of SR \cite{Book,Levy,Guerra}, and with fundamental implications for physics. This dynamical 3-space provides an explanation for the success of the SR Einstein formalism.  Indeed there is a mapping from the physical Lorentzian space and time coordinates to the non-physical spacetime coordinates of the Einstein formalism - but it is a singular map in that it removes the 3-space velocity wrt an observer. The Einstein formalism transfers dynamical effects, such as length contractions and clock slowing effects, to the metric structure of the spacetime manifold, where these effects then appear to be merely perspective effects for different observers. For this reason the Einstein formalism has been very confusing.  Developing the Lorentzian interpretation has lead to a new account of gravity, which turns out to be a quantum effect \cite{Schrod}, and of cosmology  \cite{Review,Book,Hubble, QC}, doing away with the need for dark matter and dark energy. So the discovery of the flyby anomaly links this effect to various phenomena in the emerging new physics.

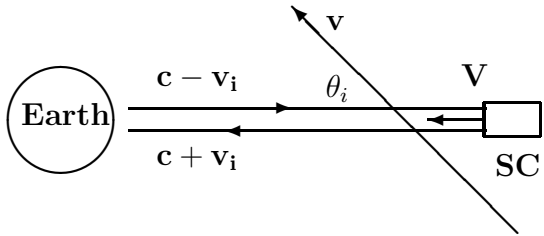
\begin{figure}
\vspace{27mm}
\hspace{40mm}
\setlength{\unitlength}{1.5mm}
\hspace{0mm}\begin{picture}(0,0)
\thicklines
\put(6.5,5){\circle{20.0}}
\put(3,4.5){\bf Earth}
\put(12.5,6){\line(1,0){31.7}}
\put(12.5,4){\line(1,0){31.7}}
\put(24,6.0){\vector(1,0){3}}
\put(24,4.0){\vector(-1,0){3}}

\put(44,3.5){\line(1,0){5}}
\put(44,6.5){\line(1,0){5}}
\put(49,3.5){\line(0,1){3}}
\put(44,3.5){\line(0,1){3}}
\put(45,0.0){\bf SC}

\put(47,-5){\vector(-1,1){20}} 

\put(44,5){\vector(-1,0){5}} 

\put(42,8){$\bf   V$} 

\put(15,8){$\bf    c-v_i$}

\put(15,1){$\bf    c+v_i$}
\put(30,13){$\bf    v$}
\put(30,7){$\theta_i$}

\end{picture}
\vspace{10mm}
\caption{ \small {Asymptotic flyby configuration in Earth frame-of-reference, with spacecraft (SC) approaching Earth with velocity $\bf V$. The departing asymptotic velocity will have a different direction but the same speed, as no force other than conventional Newtonian gravity is assumed to be acting upon the SC. The Dynamical 3-space velocity is ${\bf v}({\bf r},t)$, which causes the outward EM beam  to have speed $c-v_i$, and inward speed  $c+v_i$, where $v_i=v\cos(\theta_i)$, with $\theta_i$ the angle between $\bf v$ and $\bf V$. }}

 \label{fig:doppler}
\end{figure}

\section{Absolute Motion and Flyby Doppler Shifts}

The motion of  spacecraft relative to the earth are measured by observing the direction   and doppler shift of the   transponded RF transmissions.  As shown herein  this data gives another  technique to  determine the speed and direction of the dynamical 3-space,   manifested as a light speed anisotropy. Up to now the repeated detection of the anisotropy of the speed of light has been ignored in analysing the doppler shift data, causing the long-standing anomalies in the analysis   \cite{And2008,ref1,ref3,ref4,ref5,ref6}.

\begin{table*}
\hspace{-5mm}
{\footnotesize
\begin{tabular}{lcccccc}
\hline\hline
Parameter & GLL-I  & GLL-II & NEAR & Cassini & Rosetta & M'GER \\
\hline
Date  &  Dec 8, 1990 &  Dec 8, 1992 &  Jan 23, 1998 & Aug 18, 1999  & Mar 4, 2005  & Aug 2, 2005   \\

$V_\infty$ km/s &  8.949 & 8.877  &  6.851 & 16.010 & 3.863  &  4.056  \\
$\alpha_i$ deg  & 266.76  & 219.35  &  261.17 & 334.31  & 346.12  &  292.61 \\
$\delta_i$ deg  & -12.52  & -34.26  &-20.76   &  -12.92 & -2.81  &  31.44 \\
$\alpha_o$ deg  &219.97   & 174.35  &  183.49 & 352.54  &   246.51&   227.17 \\
$\delta_o$ deg  &-34.15& -4.87  &  -71.96 & -20.7 & -34.29  & -31.92  \\

\hline
$\alpha_v$ deg(hrs)  &108.8(7.25)   & 129.0(8.6)  &  108.8(7.25) & 45.0 (3.0)  &   130.5(8.7)&   168.0 (11.2)\\
$\delta_v$ deg  &-76& -80  &  -76 & -75 & -80 & -85  \\
$v$ km/s  &420& 420  &  450 & 420& 420 &420 \\
$\theta_i$ deg  &90.5& 56.4 & 81.8 & 72.6 & 95.3  &124.2 \\

$\theta_f$ deg  &61.8& 78.2& 19.6& 76.0 & 60.5 & 55.6  \\

\hline
(O) $\Delta V_\infty$ mm/s  & 3.92$\pm$0.3  &  -4.6$\pm$1.0 &  13.46$\pm$0.01 & -2$\pm$1  &  1.80$\pm$0.03 & 0.02$\pm$0.01  \\
(P) $\Delta V_\infty$ mm/s  & 3.92$\pm$0.1  & -4.60$\pm$0.6 &13.40$\pm$0.1 & -0.99$\pm$1.0  &  1.77$\pm$0.3 & 0.025$\pm$0.03  \\ \hline
\hline

\end{tabular}}

\caption{\small Earth flyby parameters from \cite{And2008} for spacecraft Galileo (GLL: flybys I and II), NEAR, Cassini, Rosetta and MESSENGER (M'GER).
$V_\infty$ is the average osculating hyperbolic asymptotic speed, $\alpha$ and $\delta$ are the right ascension and declination of the incoming (i) and outgoing (o) osculating asymptotic velocity vectors, and (O) $\Delta V_\infty$ is the  putative ``excess speed"  anomaly deduced by assuming that the speed of light is isotropic in modeling the doppler shifts, as in (\ref{eqn:olddoppler}).   The  observed (O) $\Delta V_\infty$  values are  from \cite{And2008}, and after correcting for atmospheric drag in the case of GLL-II, and thruster burn in the case of Cassini.  (P) $\Delta V_\infty$   is the predicted  ``excess speed", using (\ref{eqn:anomaly}),   taking account of the known light speed anisotropy and its effect upon the doppler shifts, using $\alpha_v$ and $\delta_v$ as the right ascension and declination of the 3-space flow velocity, having speed $v$, which has been taken to be 420km/s in all cases, except for NEAR, see Fig.\ref{fig:CelestialPlot}.  The $\pm$ values on  (P) $\Delta V_\infty$ indicate changes caused by changing the declination by 5\% - a sensitivity indicator.  The angles $\theta_i$ and $\theta_f$ between the 3-space velocity  and the asymptotic initial/final SV velocity $V$ are also given.  The observed doppler effect is in exceptional  agreement with the predictions using (\ref{eqn:anomaly})  and the previously measured 3-space velocity.   The flyby doppler shift is thus a new technique to accurately measure the dynamical 3-space velocity vector, albeit retrospectively from existing data. Note: By fine tuning the   $\alpha_v$ and $\delta_v$ values  for each flyby a perfect fit to the observed  (O) $\Delta V_\infty$ is possible. But here we have taken, for simplicity, the same values for GLL-I and NEAR.
}
\label{tab:table1}
\end{table*}

In the earth frame of reference, see Fig.\ref{fig:doppler}, let the transmitted signal from earth  have frequency $f$, then the corresponding outgoing wavelength is $\lambda_o=(c-v_i)/f$, where $v_i=v\cos(\theta_i)$.  This signal is received by the SC to have period $T_c=\lambda_o/(c-v_i+V)$ or frequency $f_c=(c-v_i+V)/\lambda_o$. The signal is re-transmitted with the same frequency, and so has wavelength $\lambda_i=(c+v_i-V)/f_c$, and is detected at earth with frequency $f_i=(c+v_i)/\lambda_i$.  Then overall we obtain\footnote{In practice the analysis is more complex as is the doppler shift technology. The analysis herein is sufficient to isolate and quantify the light-speed anisotropy effect. } 
\begin{equation}
f_i=\frac{c+v_i}{c+v_i-V}.\frac{c-v_i+V}{c-v_i}f
\label{eqn:frequency}
\end{equation}
Ignoring the projected 3-space velocity $ v_i$, that is, assuming that the speed of light is invariant as per the usual literal interpretation of the Einstein 1905 light speed postulate,   we obtain instead
\begin{equation}
f_i=\frac{c+V}{c-V}f
\label{eqn:oldfrequency}
\end{equation}
The use of  (\ref{eqn:oldfrequency}) instead of  (\ref{eqn:frequency})  is the origin of the putative anomalies.  The doppler shift data is usually presented in the form of speed anomalies.  Expanding  (\ref{eqn:oldfrequency})
we obtain
\begin{equation}
\frac{\Delta f_i}{f}=\frac{ f_i-f}{f}=\frac{2V}{c}+..
\label{eqn:olddopplera}
\end{equation}
From the observed doppler shift data acquired during a flyby, and then  best fitting the trajectory,  the asymptotic hyperbolic speeds   $V_{i\infty}$ and  $V_{f\infty}$ are inferred,  but incorrectly so, as in  \cite{And2008}. These inferred asymptotic speeds may be related to an inferred asymptotic doppler shift:
\begin{equation}
\frac{\Delta f_i}{f}=\frac{ f_i-f}{f}=\frac{2V_{i\infty}}{c}+..
\label{eqn:olddoppler}
\end{equation}
However expanding (\ref{eqn:frequency}) we obtain, for the same doppler shift\footnote{We ignore terms of order $vV/c^2$ within the parentheses, as in practice they are smaller than the $v^2/c^2$ terms.}
\begin{equation}
V_{i\infty}\equiv\frac{\Delta f_i}{f}.\frac{c}{2}=\frac{ f_i-f}{f}.\frac{c}{2}=\left(1+\frac{v_i^2}{c^2}\right)V+ .... 
\label{eqn:newspeedi}
\end{equation}
where $V$ is the actual asymptotic speed. 
Similarly after the flyby we obtain
\begin{equation}
V_{f\infty}\equiv -\frac{\Delta f_f}{f}.\frac{c}{2}=-\frac{ f_f-f}{f}.\frac{c}{2}=\left(1+\frac{v_f^2}{c^2}\right)V+ .... 
\label{eqn:newspeedf}
\end{equation}
and we see that the ``asymptotic" speeds   $V_{i\infty}$ and $V_{f\infty}$  must differ, as indeed first noted in the data by \cite{ref3}.  We then obtain the expression for the so-called flyby anomaly
\begin{eqnarray}
\Delta V_\infty &=& V_{f\infty}- V_{i\infty} =\frac{v_f^2-v_i^2}{c^2}V+..\nonumber \\
&=&\frac{v^2}{c^2}(\cos(\theta_f)^2-\cos(\theta_i)^2)V_\infty+..
\label{eqn:anomaly}
\end{eqnarray}
where here $V \approx V_\infty$ to sufficient accuracy, where $ V_\infty$  is the average of  $V_{i\infty}$ and $V_{f\infty}$, 
The existing data on $\bf v$ permits {\it ab initio} predictions for  $\Delta V_\infty$, and as well a separate least-squares-fit to the individual flybys permits the determination of the  direction of the 3-space velocity, relative to the earth, during each flyby, given a speed $v$, see Fig.\ref{fig:CelestialPlot}. These results are all remarkably consistent with  the data from the 11 previous  laboratory experiments that studied  $\bf v$. Note that whether the 3-space velocity is $+\bf v$ or $-\bf v$ is not material to the analysis herein, as the flyby effect is 2nd order in $v$.

\begin{figure*}
\vspace{-40mm}
\hspace{20mm}\includegraphics[scale=0.7]{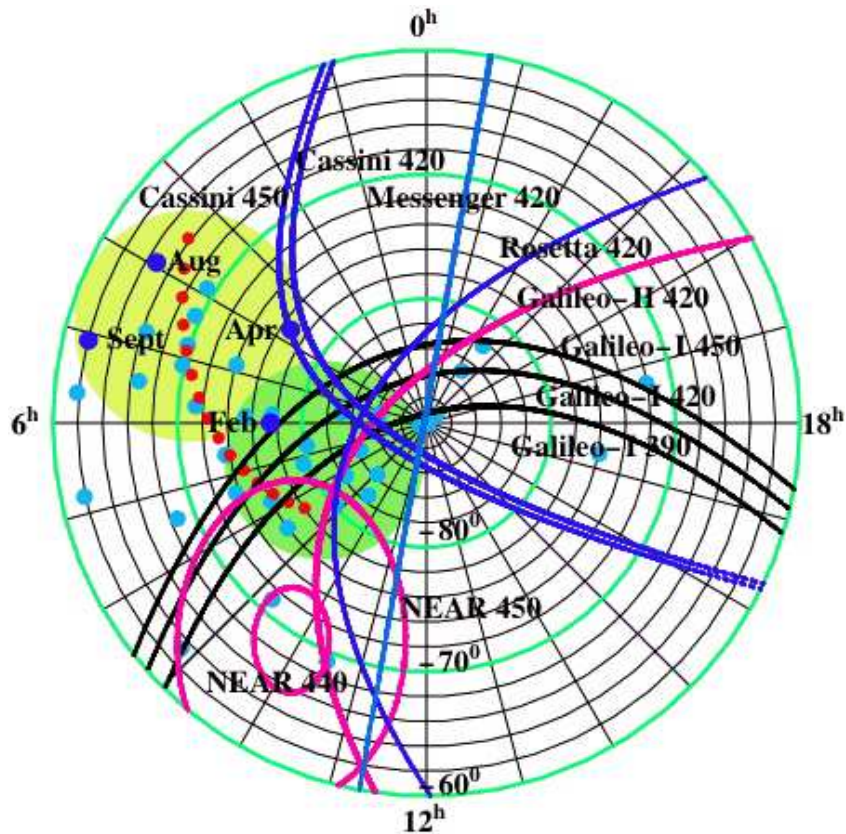}
\vspace{-3mm}
\caption{  \small Southern celestial sphere with RA and Dec shown. The 4 dark blue points show the consolidated results from the Miller gas-mode Michelson interferometer \cite{Miller} for four months in 1925/1926, from \cite{Book}.  The sequence of red points show the running daily average RA and Dec trend line,  as determined from the optical fiber interferometer data in \cite{CahillOF2},  for every 5 days, beginning September 22, 2007.  The light-blue scattered points show the RA and Dec for individual days from the same experiment, and    show significant turbulence/wave effects.  The curved plots show iso-speed  $\Delta V_{\infty}$ `anomalies':  for example for  $v = 420$km/s the RA and Dec  of ${\bf v}$ for the Galileo-I flyby must lie somewhere along the ``Galileo-I 420" curve.  The available spacecraft data  in Table 1, from \cite{And2008}, does not permit a  determination of a unique $ {\bf v}$  during that flyby.  In the case of` ``Galileo-I" the curves are also shown for $420\pm30$km/s, showing the sensitivity to the range of speeds discovered in laboratory experiments. We see that the  ``Galileo-I" December  flyby has possible directions that overlap with the December data from the optical fiber interferometer, although that does not exclude other directions, as the wave effects are known to be large.  In the case of  NEAR we must have $v \geq 440$km/s otherwise no fit to the NEAR  $\Delta V_{\infty}$ is possible.  This demonstrates a fluctuation in $v$ of at least $+20$km/s on that flyby day.  This plot shows the remarkable concordance in speed and direction from the laboratory techniques with the flyby technique in measuring ${\bf v}$, and its fluctuation characteristics.  The upper-left coloured  disk  (radius=8$^\circ$) shows concordance for September/August interferometer data and Cassini flyby data ( MESSENGER data is outside this region - but has very small $\Delta V_{\infty}$ and large uncertainty), and  the same, lower disk, for December /January/February/March data (radius=6$^\circ$).  The moving concordance effect is undertsood to be caused by the earth's orbit about the Sun, while the yearly average  of $420\pm30$km/s is a galaxy related velocity.  Directions for each flyby ${\bf v}$ were selected and used in Table 1. }
  \label{fig:CelestialPlot}
\end{figure*}

\section{Earth Flyby Data Analaysis}

Eqn.(\ref{eqn:anomaly}) permits the speed anomaly to be predicted as the direction and speed $v$ of the dynamical 3-space is known, as shown in Fig.\ref{fig:CelestialPlot}. The first determination of its direction was reported by Miller \cite{Miller} in 1933, and based on extensive observations during 1925/1926 at Mt.Wilson, California, using a large gas-mode Michelson interferometer.  These observations confirmed the previous non-null observations by Michelson and Morley \cite{MM} in 1887. The general characteristics of   ${\bf v}({\bf r},t)$ are now known following the detailed analysis  of the experiments noted above, namely its average speed,  and removing the earth orbit effect, is  some 420$ \pm $30km/s, from direction right ascension $\alpha_v=5 \pm 2^{hr}$, declination $\delta_v=70 \pm 10^\circ$S -  the center point of the Miller data in Fig.\ref{fig:CelestialPlot}, together with large wave/turbulence effects, as illustrated in Fig.\ref{fig:SeptPlot}. 
Miller's original calibration technique for the interferometer  turned out to be invalid \cite{Book}, and his speed of approximately 208km/s was recomputed to be 420 $\pm$30km/s in \cite{MMC,Book},  and the value of 420km/s is used here as shown in Table 1.  The direction of $\bf v$ varies throughout the year due to the earth-orbit effect and low frequency wave effects.  A more recent determination of the direction was reported in \cite{CahillOF2} using an optical-fiber version of the Michelson interferometer, and shown also in Fig.\ref{fig:CelestialPlot} by the trend line and data from individual days.  Directions appropriate to the date of each flyby were approximately determined from Fig.\ref{fig:CelestialPlot}.

The SC data in Table 1 shows the values of $V_\infty$ and $\Delta V_\infty$ after determining the osculating hyperbolic trajectory, as discussed in \cite{And2008}, as well as the right ascension and declination of the asymptotic SC velocity vectors $ {\bf V}_{i\infty}$ and ${\bf V}_{f\infty}$.  In computing the predicted  speed `anomaly' $\Delta V_\infty$ using  (\ref{eqn:anomaly}) it is only necessary to compute the angles $\theta_i$ and $\theta_f$ between the dynamical 3-space velocity vector and these  SC incoming and outgoing  asymptotic velocities, respectively, as we assume here that $|v|=420$kms, except for NEAR as discussed in Fig.\ref{fig:CelestialPlot} caption. So these predictions are essentially {\it ab initio} in that we are using 3-space velocities that are reasonably well known from laboratory experiments. The observed doppler effects are in exceptional  agreement with the predictions using (\ref{eqn:anomaly})  and the previously measured 3-space velocity.  The flyby anomaly is thus a new technique to accurately measure the dynamical 3-space velocity vector, albeit retrospectively from existing data.

\begin{figure}[t]
\hspace{20mm}\includegraphics[scale=1.5]{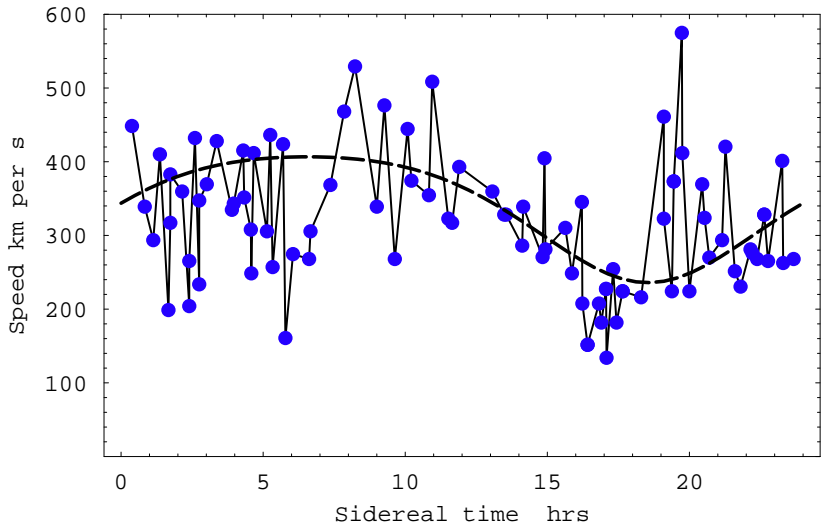}
\caption{\small {
Speeds $v_P$, of the 3-space velocity ${\bf v}$  projected onto the horizontal plane of the  Miller gas-mode Michelson interferometer,  plotted against local sidereal time in hours,  for a composite day, with data  collected over a number of days in September 1925.    The data shows considerable fluctuations, from hour to hour, and also day to day, as this is a composite day.  The dashed curve shows the non-fluctuating  best-fit variation over one day, as the earth rotates, causing the projection onto  the plane of the interferometer of the  velocity of the average direction of the space flow to change.  The maximum projected speed of the curve is 417km/s, and the  min/max occur at approximately 5hrs and 17hrs sidereal time (right ascension); see Fig.\ref{fig:CelestialPlot} for September.  Analysing Miller's extensive data set from 1925/26 gives average speed, after removing earth orbit effect, of 420$\pm$30 km/s, and the directions for each month shown in Fig.\ref{fig:CelestialPlot}. }}
\label{fig:SeptPlot}\end{figure}

\section{New Gravitational Waves}

Light-speed anisotropy experiments have revealed that a dynamical 3-space exists, with the speed of light being $c$, in vacuum, only wrt to this space: observers in motion `through' this 3-space detect that the speed of light is in general different from $c$, and is different in different directions.  The dynamical equations for this 3-space are now known and involve a velocity field ${\bf v}({\bf r},t)$, but where only relative velocities are observable locally  - the coordinates ${\bf r}$ are relative to a non-physical mathematical embedding space. These dynamical equations involve Newton's gravitational constant $G$ and the fine structure constant $\alpha$.  The discovery of this dynamical 3-space then required a generalisation of the Maxwell, Schr\"{o}dinger and Dirac equations.     The wave effects already detected correspond to fluctuations in the 3-space velocity field ${\bf v}({\bf r},t)$, so they are really 3-space turbulence or wave effects.  However they are better known, if somewhat inappropriately, as `gravitational waves' or `ripples' in `spacetime'.  Because the 3-space dynamics gives a deeper understanding of the spacetime formalism we now know that the metric of the induced spacetime, merely a mathematical construct having no ontological significance, is related to ${\bf v}({\bf r},t)$ according to \cite{Review,Book,QC}
\begin{equation}
ds^2=dt^2 -(d{\bf r}-{\bf v}({\bf r},t)dt)^2/c^2
=g_{\mu\nu}dx^{\mu}dx^\nu
\label{eqn:Eqn1}\end{equation}
The gravitational acceleration of matter,  a quantum effect, and of the structural patterns  characterising  the 3-space, are given by \cite{Review,Schrod}
\begin{equation}
{\bf g}=\frac{\partial {\bf v}}{\partial t}+({\bf v}.\nabla ){\bf v}
\label{eqn:acceln}
\end{equation}
and so fluctuations in  ${\bf v}({\bf r},t)$ may or may not manifest as a gravitational acceleration.  The flyby technique assumes that the SC trajectories are not affected - only the light speed anisotropy is significant. 
 The magnitude of this turbulence depends on the timing resolution of each particular experiment, and was characterised to be sub-mHz in frequency by Cahill and Stokes\cite{CahillCoax}.   Here we have only used asymptotic osculating hyperbolic trajectory data from \cite{And2008}.   Nevertheless even this data suggests the presence of wave effects. For example the NEAR data requires a speed in excess of 440km/s, and probably closer to 450km/s, whereas the other flybys are consistent with the average of 420km/s from laboratory experiments. So here we see flyby evidence of fluctuations in the speed $v$.

 Data exists for each full flyby, and analysis of that data using the new doppler shift theory will permit  the study and characterisation of the 3-space wave turbulence during each flyby: essentially the flybys act as gravitational wave detectors.  These gravitational waves are much larger than predicted by general relativity, and have different properties.  

\section{Pioneer10/11 Anomalies}
The Pioneer 10//11 spacecraft have been exploring the outer solar system since 1972/73.  The spacecraft have followed escape hyperbolic orbits near the plane of the ecliptic, after earlier planet flybys. The doppler shift data,  using (\ref{eqn:oldfrequency}), have revealed an unexplained anomaly beyond 10 AU \cite{PioneerDoppler}.   This  manifests as an  unmodelled increasing blue shift  $\displaystyle{ \frac{d}{dt}(\frac{\Delta f}{f})}=(2.92\pm0.44)\times10^{-18}s/s^2$, corresponding to a constant inward sun-directed acceleration of  $a=\displaystyle{\frac{dV}{dt}}=(8.74\pm1.33)\times10^{-8}$ cm/s$^2$, averaged  from Pioneer 10 and  Pioneer 11 data.  However the doppler-shift data from these spacecraft has been interpreted using  (\ref{eqn:oldfrequency}), instead of (\ref{eqn:frequency}), in determining the speed, which in turn affects the distance data.  Essentially this implies that the spacecraft are attributed with a speed that is too large by $\displaystyle{\frac{v^2}{c^2}V_D}$, where $V_D$ is the speed determined using  (\ref{eqn:oldfrequency}). This then implies that the spacecraft are actually  closer to the sun by the distance $\displaystyle{\frac{v^2}{c^2}R_D}$,  where $R_D$ is the distance determined using  (\ref{eqn:oldfrequency}). This will then result in a computed  spurious inward acceleration,  because the gravitational pull of the sun is actually larger than modelled, for distance $R_D$.  However this correction to the doppler-shift analysis appears not to be  large enough to explain the above mention acceleration anomaly. Nevertheless re-analysis of the Pioneer 10/11 data should be undertaken using  (\ref{eqn:frequency}), particularly as the anomaly began after a planetary flyby.

\section{Conclusions}

The spacecraft earth flyby anomalies have been resolved. Rather than actual relative changes in the asymptotic inward and outward speeds, which would have perhaps required the invention of a new force, they are instead direct manifestations of the anisotropy of the speed of light, with the earth having a speed of some 420 $\pm$30km/s relative to a dynamical 3-space, a result consistent with previous determinations using laboratory experiments,  and dating back to the Michelson-Morley 1887  experiment, as recently reanalysed \cite{MMCK,MMC,Review}.  The flyby data also reveals, yet again, that the 3-space velocity fluctuates in direction and speed, and with results also consistent with laboratory experiments.  Hence we see a remarkable concordance between three different laboratory techniques, and the newly recognised flyby technique.  The existing flyby data  can now be re-analysed to give a  detailed charaterisation of these gravitational waves.  The detection of the 3-space velocity gives a new astronomical window on the galaxy, as the observed speeds are those relevant to galactic dynamics.  The dynamical 3-space velocity effect also produces very small vorticity effects when passing the earth, and these  are predicted to produce observable effects on the GP-B gyroscope precessions \cite{GPB}.

A  special acknowledgement   to all the researchers who noted and analysed the spacecraft anomalies, providing the excellent data set used herein.  Thanks also to Tom Goodey for encouraging me to examine these anomalies.

\end{document}